\documentclass{fmj2010}
\usepackage{graphicx}

\def\fermilat{\textit{Fermi}/LAT}

\begin{document}
   \title{Radio Flaring Activity of 3C~345 and its Connection to
$\gamma$-Ray
Emission}

   \author{F.~K. Schinzel\inst{1,4}\thanks{Member of the International
Max-Planck Research School (IMPRS) for Astronomy and Astrophysics at the
Universities of Bonn and Cologne.}
          \and
          A.~P. Lobanov\inst{1}
	  \and
	  S.~G. Jorstad\inst{3}
	  \and
	  A.~P. Marscher\inst{3}
	  \and
	  G.~B. Taylor\inst{2,4}\thanks{Also an Adjunct
Astronomer at the National Radio Astronomy Observatory.}
	  \and
	  J.~A. Zensus\inst{1,4}
          }

   \institute{Max-Planck-Institut f\"ur Radioastronomie, Auf dem H\"ugel 69,
53121 Bonn, Germany
         \and
             Department of Physics and Astronomy, University of
New Mexico, Albuquerque NM, 87131, USA
	 \and
	     Institute for Astrophysical Research, Boston University, 725
Commonwealth Avenue, Boston, MA 02215
	 \and On behalf of the \fermilat\ collaboration
             }

   \abstract{ 3C~345 is one of the archetypical active galactic
nuclei, showing structural and flux variability on parsec scales
near a compact unresolved radio core. During the last 2 years, the
source has been undergoing a period of high activity visible in the
broad spectral range, from radio through high-energy bands. We have
been monitoring parsec-scale radio emission in 3C 345 during this
period at monthly intervals, using the VLBA at 15, 24, and 43~GHz.
Our radio observations are compared with gamma-ray emission detected
by Fermi-LAT in the region including 3C~345 (1FGL J1642.5+3947). Three
distinct gamma-ray events observed in this region are associated
with the propagation of relativistic plasma condensations inside the
radio jet of 3C~345. We report on evidence for the gamma-rays to be
produced in a region of the jet of up to 40 pc (de-projected) in
extent. This suggests the synchrotron self-Compton process as
the most likely mechanism for production of gamma-rays in the source.
}

   \maketitle
%

\section{Introduction}

The quasar \object{3C~345} is one of the best studied
``superluminal'' radio sources, with its parsec-scale radio emission
monitored over the past 30 years. Substantial variability of the
optical (\cite{1984Ap.....20..461B,1990A&A...239L...9K}) and radio
(\cite{1996ASPC..110..208A, 1998A&AS..132..305T, 1999ApJ...521..509L})
emission has been observed, with a possible periodicity of 3.5--4.5
years and major flares occurring every 8--10 years. A new cycle of
such enhanced nuclear activity began in early 2008
(\cite{2009ATel.2222....1L}).

3C~345 has also been known as a prominent variable source at high
energies up to the X-ray band and only in the $\gamma$-ray regime had it not
been clearly detected (\cite{2008A&A...489..849C}), possibly
due to the low spatial resolution of previous $\gamma$-ray instruments and
a lack of space instruments during periods of high source
activity. The launch of the GLAST satellite (now \textit{Fermi})
equipped with the Large Area Telescope (LAT) survey instrument
(\cite{2009ApJ...697.1071A}) enabled continuous monitoring of
$\gamma$-ray emission originating from the vicinity of 3C~345.

This paper presents first results from an analysis of
\fermilat\ $\gamma$-ray monitoring data, combined with monthly radio
observations made at 43.2~GHz (7~mm wavelength) with very long
baseline interferometry (VLBI), using the VLBA\footnote{Very Long
Baseline Array of the National Radio Astronomy Observatory, Socorro,
USA} facility.

Throughout this paper, a flat $\Lambda$CDM cosmology is assumed, with
$H_0 = 71$\,km\,s$^{-1}$\,Mpc$^{-1}$ and $\Omega_\mathrm{M}$ = 0.27. At the
redshift $z=0.593$ (\cite{2007AJ....134..102S}) of 3C~345, this corresponds to a
luminosity distance $D_\mathrm{L} = 3.47$\,Gpc, a linear scale of 6.64 pc
mas$^{-1}$, and a proper motion scale of 1$\,$mas$\,$year$^{-1}$ corresponding
to 34.5$\, $c.

\section{Observations \& Data Processing}

\subsection{\fermilat}

The 11-month all-sky $\gamma$-ray monitoring by the \fermilat\ 
instrument lists a strong detection of a $\gamma$-ray emitter in the region
\object{1FGL~J1642.5$+$3947} (\cite{2010arXiv1002.2280T, 2010ApJ...715..429A}),
which lies in the close vicinity of the three quasars \object{3C~345}
($\sim$6.4\arcmin\ away from $\gamma$-ray localization) \object{CLASS
J1641+3935} ($\sim$17\arcmin\ away) and
\object{NRAO~512} ($\sim$31\arcmin\ away). The three month
bright AGN list mentions the source 0FGL J1641.4+3939
(\cite{2009ApJ...700..597A}) as being associated with the faint blazar CLASS
J1641+3935 with low probability. However, the 1FGL source is now more
convincingly associated with 3C~345 and not the faint CLASS source, albeit still
formally at a low-probability.

The angular resolution (at a 68\% confidence) of the \fermilat\
instrument is about $0.8^\circ$ for single photons at 1~GeV energy and
it is worse at other energies (\cite{2009ApJ...697.1071A}), thus
making it difficult to obtain a significant association with any source
in this particular field, based on the $\gamma$-ray data alone. The
statistical analysis shows that a majority of the received photons
are located closer to 3C~345 than to the other two sources. Contribution by
the other two candidates seems to be minor. A firm association of
1FGL~J1642.5$+$3947 with 3C~345 can only be established with the
inclusion of multi-wavelength observations. A detailed description of
the maximum likelihood analysis as well as a thorough discussion of
the $\gamma$-ray data will be given in a forthcoming \fermilat\
collaboration paper on the $\gamma$-ray detection and multi-wavelength
identification of 3C~345 (\cite{2010Apj_prep}).

\begin{figure}
		\centering
		\includegraphics[width=0.45\textwidth]{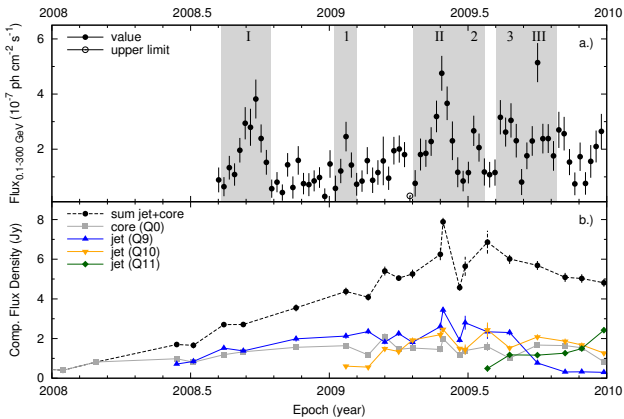}
		\caption{ \textit{a.) top}: \textit{Fermi}-LAT weekly
$\gamma$-ray lightcurve of 3C~345 for the energy range 0.1~-~300~GeV. The
integrated values have a test statistics value of at least 5, corresponding to
a $\geq 2.2 \sigma$ detection. \textit{b.) bottom:}
VLBA 7~mm component flux densities for the model-fitted VLBI core and inner jet,
represented by up to four circular Gaussians (Q0, Q11, Q10, Q9). The
component labeled Q0 is the eastern-most component (see Fig.~\ref{fig:map}) and
represents the compact ``core'' or base of the jet. The dashed line plots
the sum of the flux densities of all four components.}
		\label{fig:lightcurve}
	\end{figure}

From the available \fermilat\ data, a weekly light curve was
constructed for the first 14~months of LAT observations using the \textit{Fermi}
Science Tools software package release v9r15p2 (08/08/2009). The procedures
applied were similar to the one described in \cite{2009ApJ...700..597A} using
the current instrument response function P6\_V3, isotropic background
model v02, and Galactic diffuse background models as discussed in
\cite{2009ApJS..183...46A}. The sources Mrk~501 and 4C~+38.41 have been included
in the source model for the region of interest using a powerlaw spectral fit
with photon spectral energy indices\footnote{The photon spectral index $n$ is
defined as $F(E)\, \propto\, E^{-n}$, where F(E) is the
$\gamma$-ray photon flux as function of energy E.} of around $1.79$ and $2.66$
respectively. For the target source 3C~345, a fixed photon index of $2.46$ was
applied to obtain weekly averaged flux values from the modelfits. The
resulting 0.1--300\,GeV lightcurve of 3C~345 is shown in the top panel of
Fig.~\ref{fig:lightcurve}.

\subsection{VLBA}

Following the onset of a new period of flaring activity in 2008, we
initiated a dedicated monthly VLBA monitoring of radio emission from
3C~345 at 43.2, 23.8, and 15.4~GHz (VLBA project codes: BS193,
BS194). In this paper only the 43.2~GHz observations are discussed,
while the analysis of 15.4 and 23.8~GHz data is continued. The
observations were made with a bandwidth of 32~MHz (total recording bit
rate 256~Mbit\,s$^{-1}$). The total of 12 VLBA observations have been
completed, with about 4.5 hours at 43.2~GHz spent on 3C~345 during
each observation. Scans on \object{3C~345} were interleaved with
observations of \object{J1310+3233} (amplitude check, EVPA
calibrator), \object{J1407+2827} (D-term calibrator), and
\object{3C~279} (amplitude check, EVPA calibrator). The VLBA data were
correlated at the NRAO VLBA processor. Analysis was done with the NRAO
Astronomical Image Processing System (AIPS) and Caltech Difmap
(\cite{1995BAAS...27..903S}) software for imaging and
modeling. Corrections were applied for the parallactic angle and for
Earth orientation parameters used by the VLBA correlator. Fringe
fitting was applied to calibrate the observations for group delay and
phase rate. A summary of all the observations is presented in
Table~\ref{tab:obssummary}. The data from the first 10 epochs of the
projects BS193 and BS194 are complemented with 14 VLBA observations
from the blazar monitoring program of Marscher et al. (VLBA project
codes BM256, BM303, S1136) available
online\footnote{http://www.bu.edu/blazars/VLBAproject.html}. The
combined data (see Table~\ref{tab:obssummary}) cover the period from
January 2008 to December 2009, with observations spaced roughly at
monthly intervals.

\begin{table}
 		\caption{Summary of 43~GHz VLBA Observations.}
		\label{tab:obssummary}
		\begin{center}
		\small
		\begin{tabular}{c c c c c}
 			\hline\hline
			Date & $S_\mathrm{tot}$ & $D$ & Beam (bpa) &  Ref. \\
			     & [Jy] &   & [mas]x[mas] (deg)  & \\
			\hline
			2008-01-17 & 1.90 & 2300 & 0.31$\times$0.19 (-27) &
1\\
			2008-02-29 & 2.02 & 1600 & 0.37$\times$0.21 (-29) &
1\\
			2008-06-12 & 2.44 & 2100 & 0.38$\times$0.16 (-28)
 & 1\\
			2008-07-06 & 2.23 & 800  & 0.33$\times$0.15 (-17) &
1\\
			2008-08-16 & 3.78 & 2700 & 0.41$\times$0.19 (-30) &
1\\
			2008-09-10 & 3.67 & 4300 & 0.37$\times$0.19 (-33) &
1\\
			2008-11-16 & 4.43 & 300 & 0.37$\times$0.33 (-1.5) &
1\\
			2008-12-21$^\dag$ & 2.91  & 5500 & 0.39$\times$0.17
(-17) & 1\\
			2009-01-24 & 5.04 & 8900 & 0.31$\times$0.17 (-21) &
1\\
			2009-02-19$^\dag$ & 3.55 & 2200 & 0.37$\times$0.20
(-19) & 2\\
			2009-02-22 & 4.63 & 8800 & 0.35$\times$0.15 (-19) &
1\\
			2009-03-16 & 6.01 & 2400 & 0.43$\times$0.30 (9.3) & 2\\
			2009-04-01 & 5.67 & 8000 & 0.33$\times$0.16 (-19) &
1\\
			2009-04-21 & 5.78  & 2600 & 0.33$\times$0.22 (-17) &
2\\
			2009-05-27 & 7.00 & 3500 & 0.33$\times$0.18 (-15) &
2\\
			2009-05-30 & 8.65 & 6900 & 0.32$\times$0.16 (-20)
& 1\\
			2009-06-21 & 5.04 & 6500 & 0.28$\times$0.16 (-11) &
1\\
			2009-06-29 & 6.43 & 2600 & 0.29$\times$0.16 (-12) &
2\\
			2009-07-27 & 7.63 & 2900 & 0.38$\times$0.16 (-31) &
1\\
			2009-08-26 & 6.58 & 2000 & 0.32$\times$0.17 (-25) &
2\\ 
			2009-10-01 & 6.67 & 2300 & 0.22$\times$0.18 (-22) &
2\\
			2009-11-07 & 6.38 & 1400 & 0.32$\times$0.16 (-15) &
2\\
			2009-11-30 & 5.53 & 2000 & 0.29$\times$0.16 (-17) &
2\\
			2009-12-28 & 4.95 & 2400 & 0.31$\times$0.16 (-8.1) &
2\\
			\hline
		\end{tabular}\\[12pt]
		\end{center}
		\begin{flushleft}
		\small
		{\bf Notes:} $S_\mathrm{tot}$ -- total flux density recovered in
VLBA image;
		$D$ -- dynamic range measured as a ratio of the image peak flux
density to the r.m.s. noise;
		Beam (bpa) -- beam size, major axis vs minor axis with position
angle of ellipse in parenthesis;
		References: 1 -- blazar monitoring Marscher et al. (VLBA project
code BM256, BM303,
S1136); 2 -- dedicated monitoring (VLBA project codes BS193, BS194).
		$\dag$ -- not used for the flux density analysis due to gain
calibration problems and bad weather conditions at some of the VLBA antennas.
		\end{flushleft}
\end{table}

\begin{figure}
		\centering
		\includegraphics[width=0.45\textwidth]{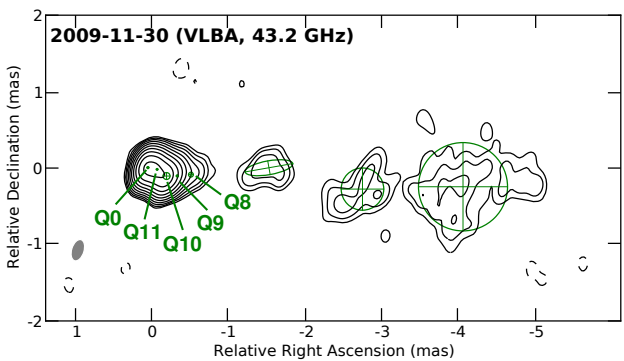}
		\caption{VLBA image of total intensity of 3C~345 at 43.2~GHz
made from observations on Nov.~30, 2009. Open ellipses show the FWHM of eight
Gaussian components used to fit the structure observed. The shaded ellipse
represents the FWHM of the restoring beam. The image peak flux density is
2.1~Jy~beam$^{-1}$ and the r.m.s. noise is 1~mJy beam$^{-1}$. The contour 
levels are (-0.15, 0.15, 0.3, 0.6, 1.2, 2.4, 4.8, 9.6, 19, 38, 77)~\%
of the peak flux density.
}
		\label{fig:map}
	\end{figure}

The brightness distribution of the radio emission was model-fitted by
multiple Gaussian components, providing positions, flux densities and
sizes of distinct emitting regions in the jet. Fig.~\ref{fig:map} illustrates
the observed radio structure and its Gaussian model fit representation. Compact
emission in the nuclear region ($\leq0.15$~mas from the VLBI core) was modelled
using two circular Gaussians providing the optimal ratio of $\chi^2$ to the
number of model parameters for all epochs.

We interpret the eastern-most component, hereafter labeled Q0,
as the base (or ``core'') of the radio jet at 43.2~GHz, whereas the other
features can signify perturbations or shocks developing in the jet. Locations
and proper motions of other jet features are then determined with respect to Q0.

\section{Results}

\subsection{Evolution of the radio emission in the nuclear region}

A new moving emission region, labeled Q9, was first detected with the
VLBA observation on June 16, 2008, followed by detections of another
new component on January 24, 2009 (Q10) and a third one on July 27,
2009 (Q11). In the following, the components Q9, Q10, and Q11 observed
within a distance of $\leq 0.3$~mas from the core Q0 are
referred to as the ``jet''.

The flux density evolution of the jet is plotted in the lower part of
Fig.~\ref{fig:lightcurve} together with the flux densities of the core
and the sum of jet and core. During 2009, the jet (average flux
density: 3.6~Jy) was stronger than the core (average flux density:
1.5~Jy) by a factor of 2.4.

Relative positional offsets of components Q10, Q9, and Q8 are
measured with respect to the base of the jet at component Q0. All three
features follow a similar trajectory; however, curiously, the recently detected
Q11 shows a north-ward offset of 0.025~mas ($\sim$1/7th of the
beamsize) in its trajectory, compared to the previous features. This should be
verified by continued radio monitoring of the jet. 

All newly ejected jet features show similar apparent acceleration from about 2 -
10~c over a distance of 0.2~mas. 

\subsection{$\gamma$-Ray Emission}

The weekly binned light curve plotted in Fig.~\ref{fig:lightcurve}a.,
shows six distinct events above
$2.4\cdot10^{-7}\,\mathrm{ph}\,\mathrm{cm}^{-2}\,\mathrm{s}^{-1}$.

The six events are split into two sub-categories according to their
duration. Long events, with durations of 40--60 days, are labeled with roman
capitals; shorter events, with durations of 20--35 days, are labeled with arabic
numbers.
These sub-categories can also be distinguished by their peak fluxes: the long
events had peak flux values of $\left(3.86 - 5.19\right)
\cdot10^{-7}\,\mathrm{ph}\,\mathrm{cm}^{-2}\,\mathrm{s}^{-1}$, whereas
short events ended up with a lower range of $\left(2.48 - 3.19\right)
\cdot10^{-7}\,\mathrm{ph}\, \mathrm{cm}^{-2}\, \mathrm{s}^{-1}$.

	Flare III has been reported on previously in the form of two Astronomers
Telegrams by the \fermilat\ (\cite{2009ATel.2226....1R}) and GASP collaborations
(\cite{2009ATel.2222....1L}), reporting on
mm-wavelength (230 GHz SMA), optical
(R-Band) and $\gamma$-ray (GeV) activity. 

\subsection{Radio/$\gamma$-ray Correlation}

	A first look at the $\gamma$-ray light curve (Fig.~\ref{fig:lightcurve}
a) gives the impression of a rising underlying trend in the $\gamma$-ray
emission similar to the one observed in radio. 

\begin{figure}
		\centering
		\includegraphics[width=0.45\textwidth]{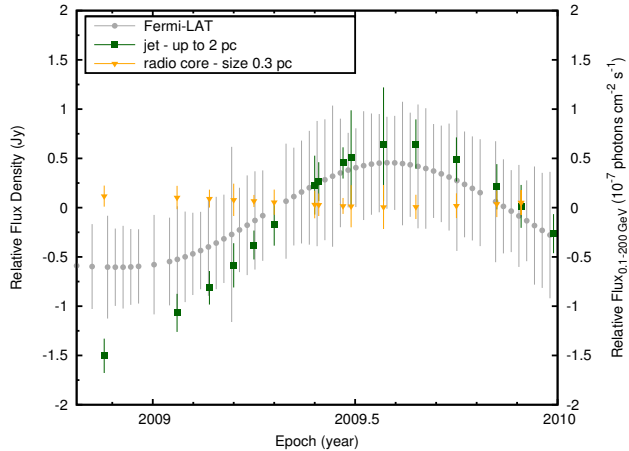}
		\caption{Long-term trends of the radio jet, 
radio core flux densities and the $\gamma$-ray flux relative to their respective
mean values (jet: 3.6~Jy, core: 1.5~Jy, $\gamma$: $1.9\cdot
10^{-7}$~ph~cm$^{-2}$~s$^{-1}$). The trend has been obtained using splines
with 0.5 year bins. $\gamma$-ray and radio light curves between 2009.8 and
2010.0 have been re-scaled. The 0 value represents the mean value of
3.60~Jy for the radio flux density of the jet of an apparent size of
$\leq 2$~pc (excluding the core), 1.5~Jy for the flux density of the core of
an apparent size of $\sim 0.3$~pc and $1.9\cdot 10^{-7}$~ph~cm$^{-2}$~s$^{-1}$
for $\gamma$-rays.}
		\label{fig:detrend}
	\end{figure}

	The radio and $\gamma$-ray light curves were ``de-trended'' using
cubic spline interpolations with 0.5~year bins
(\cite{1992nrfa.book.....P}).  Remarkably, as shown in
Fig~\ref{fig:detrend}, the jet matches the long-term trend of
the $\gamma$-rays. A mismatch of the first 0.5~year results from the absence of
$\gamma$-ray data before 2008.6 and a secondary weaker
$\gamma$-ray flare after 2008.7 (see Fig.~\ref{fig:lightcurve}) which
contributes to the flux level at 2008.8 causing a slightly higher
starting flux. The core showed no to little variation in flux density over
this time period, with an average value around 1.5~Jy. 

	A discrete correlation as described by \cite{1988ApJ...333..646E} was
applied to the re-scaled with their respective average values and de-trended
light curves with the obtained cubic splines fits. No strong correlation is
evident, either for the jet or for the core, after de-trending. However, the
jet
shows a weak correlation with a coefficient of $0.6\pm0.3$ at timelag
$3\pm15$~days and a weak anti-correlation in the following bin of $-0.6\pm0.3$.

Sparse sampling of the radio data makes it difficult to obtain firm
specific localizations of individual events in the radio
jet. Nonetheless, the similarity of the long-term trends observed in
the lightcurves provides good evidence for correlated underlying
emission between the radio flux density of the jet and the
$\gamma$-ray flux.

\section{Summary \& Conclusions}

We have found a correspondence between the long-term \textit{Fermi}-LAT
lightcurve of the region 1FGL~J1642.5$+$3947 and the radio emission of 3C~345,
establishing the detection of $\gamma$-ray emission from 3C~345 by comparison of
radio and $\gamma$-ray variability.

More importantly, we find $\gamma$-ray emission to be related to the
pc-scale jet of the source of up to 2~pc ($\sim$~40~pc de-projected adopting a
viewing angle of $\theta \sim $~2.7$^\circ$; \cite{2005AJ....130.1418J}). We
have been able to trace back the ejection of new superluminally moving and
apparently accelerating features in the jet to be linked to $\gamma$-ray
production. Even though a correspondence between the jet and $\gamma$-ray
emission was found, there is no evidence for correlation between $\gamma$-ray
emission and radio core flux density at 43~GHz. The observed radio properties of
these features together with the observed $\gamma$-ray variability question
existing jet models and suggest the synchrotron self-Compton (SSC) process as
the most likely mechanism driving the production of $\gamma$-ray photons in the
source. This conclusion is supported by the spectral energy distribution of
3C~345 being compatible with SSC exhibiting an IC to synchrotron peak ratio of
only $\sim$5 (for the Oct. 2009 flare). 

Continued monitoring and more densely sampled VLBI observations could
provide better confirmation of our results and provide an opportunity
to localize more accurately the sites of individual flares in the
jet. At this writing, the nuclear region of 3C~345 remains at high flux density
and $\gamma$-ray flux levels are still elevated. The characteristics of
the long-term activity of the source as observed over the last 30
years suggest that this continued activity may last for at least
another year, giving a unique opportunity to trace this active state.

\begin{acknowledgements}
Frank Schinzel was supported for this research through a
stipend from the International Max-Planck Research School (IMPRS) for Astronomy
and Astrophysics at the Universities of Bonn and Cologne. The research at Boston U. was funded in part by NASA Fermi Guest
Investigator Program grants NNX08AV65G, NNX08AV61G, and NNX09AT99G,
and National Science Foundation grant AST-0907893. The National Radio Astronomy Observatory is a facility of
the National Science Foundation operated under cooperative agreement by
Associated Universities, Inc. The \textit{Fermi} LAT Collaboration acknowledges generous ongoing support from
a number of agencies and institutes that have supported both the development and
the operation of the LAT as well as scientific data analysis. These include the
National Aeronautics and Space Administration and the Department of Energy in
the United States, the Commissariat `a l’Energie Atomique and the Centre
National de la Recherche Scientifique / Institut National de Physique Nucl´eaire
et de Physique des Particules in France, the Agenzia Spaziale Italiana and
the Istituto Nazionale di Fisica Nucleare in Italy, the Ministry of Education,
Culture, Sports, Science and Technology (MEXT), High Energy Accelerator Research
Organization (KEK) and Japan Aerospace Exploration Agency (JAXA) in Japan, and
the K. A. Wallenberg Foundation, the Swedish Research Council and the Swedish
National Space Board in Sweden. This workshop has been supported by the European 
Community Framework Programme 7, Advanced Radio Astronomy in Europe, 
grant agreement no.: 227290.

\end{acknowledgements}

\end{document}